\documentclass[preprint,showpacs,preprintnumbers,amsmath,amssymb]{revtex4}

\usepackage{graphicx,amsfonts,amssymb,epsfig}

\begin{document}
\title{2D XY Behavior observed in quasi-2D quantum Heisenberg antiferromagnets}
\author{F. Xiao, F. M. Woodward, and C. P. Landee}
\affiliation{Department of Physics, Clark University,
Worcester, MA 01610, USA}
\author{M. M. Turnbull}
\affiliation{Carlson School of Chemistry and Biochemistry, Clark University,
Worcester, MA 01610, USA}
\author{C. Mielke and N. Harrison}
\affiliation{Los Alamos National Laboratory,
Los Alamos, NM 87545, USA }
\author{T. Lancaster}
\author{S. J. Blundell}
\author{P. J. Baker}
\author{P. Babkevich}
\affiliation{
Oxford University Department of Physics, Clarendon Laboratory,
Parks Road, Oxford, OX1 3PU, UK}
\author{F. L. Pratt}
\affiliation{
ISIS Facility, Rutherford Appleton Laboratory, Chilton, Didcot, Oxfordshire, OX11 0QX, UK}
\date{\today}
\begin{abstract}
The magnetic properties of a new family of molecular-based quasi-two dimension $S=1/2$ Heisenberg antiferromagnets are reported. Three compounds, (Cu(pz)$_2$(ClO$_4$)$_2$, Cu(pz)$_2$(BF$_4$)$_2$, and [Cu(pz)$_2$(NO$_3$)](PF$_6$)) contain similar planes of Cu$^{2+}$ ions linked into magnetically square lattices by bridging pyrazine molecules (pz = C$_4$H$_4$N$_2$). The anions provide charge balance as well as isolation between the layers. Single crystal measurements of susceptibility and magnetization, as well as muon spin relaxation studies, reveal low ratios of N\'{e}el temperatures to  exchange strengths ($4.25 / 17.5 = 0.243$, $3.80/15.3=0.248$, and $3.05/10.8=0.282$, respectively) while the ratio of the anisotropy fields $H_A$~(kOe) to the saturation field $H_\mathrm{SAT}$~(kOe) are small ($2.6/490 = 5.3\times$10$^{-3}$, $2.4/430=5.5\times$10$^{-3}$, and $0.07/300=2.3\times$10$^{-4}$, respectively), demonstrating close approximations to a  2D Heisenberg model. The susceptibilities of ClO$_4$ and BF$_4$ show evidence of an exchange anisotropy crossover (Heisenberg to $XY$) at low temperatures; their ordering transitions are primarily driven by the $XY$ behavior with the ultimate 3D transition appearing parasitically. The PF$_6$ compound remains Heisenberg-like at all temperatures, with its transition to the  N\'{e}el state due to the interlayer interactions. Effects of field-induced anisotropy  have been observed.
\end{abstract}
\date{\today}
\maketitle

\section{Introduction}

For more than seventy years \cite{Bethe31} the study of low-dimensional magnetism has played an integral role in the understanding of phase transitions, critical behavior, and other aspects of quantum many-body physics. Highlights of this progression include the pathbreaking neutron scattering studies of the excitation spectrum of the one-dimensional (1D) S=1/2 Heisenberg antiferromagnet\cite{Endoh74, Heilmann78}, the discovery of superconductivity in doped exchange-coupled layers of Cu(II)  oxides\cite{Bednorz86} with the consequent flurry of theoretical\cite{Manousakis91} and experimental\cite{Birgeneau98} research, as well as the discovery of macroscopic quantum tunneling in high-spin nanomagnets\cite{Friedman96, Thomas96}.

Experimental studies in quantum magnetism rely on the existence of simple and well-characterized model systems. For the case of the $S=1/2$ (quantum) two-dimensional Heisenberg antiferromagnet on a square lattice (2D QHAF), the exchange must be as close to isotropic as possible and the layers must be well isolated. The appropriate Hamiltonian $\mathcal{H}$ for a {\it quasi}-2D QHAF in an applied field $H$ is
\begin{eqnarray}
\mathcal{H}&=&J \sum_{nn} \left[ S_i^xS_{j}^x+S_i^yS_{j}^y+(1-\Delta) S_{i}^{z} S_{j}^{z}\right] \notag\\
& & +J^\prime \sum_{i,i^\prime} \mathbf{S}_{i}\cdot\mathbf{S}_{i^\prime}
-g\mu_{\mathrm{B}} \mathbf{H} \cdot \sum_{j} \mathbf{S}_{j},
\label{Hamiltonian}
\end{eqnarray}
where the first summation is over nearest neighbors in the planes, the second summation links each spin to its counterparts in adjacent layers and the third includes all spins. Here $J$ is the intralayer exchange parameter, $J^\prime$ is the coupling constant between spins in adjacent layers, and $\Delta$ is the exchange anisotropy parameter. (In this Hamiltonian, a positive $J$ corresponds to antiferromagnetic exchange.) For an ideal 2D QHAF, $J^\prime=0$ and $\Delta = 0$. A full characterization of any physical realization of a model Hamiltonian must determine experimentally the values of $J$, $J^\prime$, and $\Delta$.

Little is known about the properties of the 2D QHAF in applied magnetic fields. The Zeeman term in the Hamiltonian (Eq.~(\ref{Hamiltonian})) has generally been neglected in studies of the copper oxides due to their large energy scale. Assuming the saturation field depends only on the exchange strength, the mean-field equation for $H_\mathrm{SAT}$ of an $S=1/2$ system is given by\cite{Bonner64}
\begin{equation}
H_\mathrm{SAT} = \frac{zk_BJ}{g\mu_B}
\label{hsat}
\end{equation}
where $z$ is the number of nearest neighbors and $J$ is expressed in units of Kelvin. An exchange strength of 1500~K corresponds to an $H_\mathrm{SAT}\approx 4200$~T and the usual experimental fields are minor perturbations to the Hamiltonian.

Seven years ago some of the present authors published\cite{Woodward01, Woodward02} several studies of the molecular-based quasi-2D QHAF family which have exchange constants of 6.5 and 8.5~K, respectively, and respective saturation fields of 19 and 24~T. Although these materials did have relatively large exchange ratios, $J^\prime/J\approx 0.24$, they did show the characteristic susceptibilities\cite{Woodward01, Woodward02} and heat capacities\cite{Matsumoto2000} of the 2D QHAF and revealed for the first time the characteristic upward curvature of the low-temperature magnetization of the 2D QHAF.  Preliminary reports of this discovery led to several theoretical studies of the magnetization\cite{yang97, Zhitomirsky98} and gave motivation for further investigations of the 2D QHAF in large applied fields. More recent papers\cite{Zhitomirsky99, Syljuasen02} predict an anomalous spin excitation spectrum as the external field approaches $H_\mathrm{SAT}$. In addition, it has been predicted\cite{Cuccoli_PRB_2003, Cuccoli_PRL_2003} that the application of an external field to a 2D QHAF would reduce quantum fluctuations along the field axis and gradually transform the spin anisotropy from Heisenberg toward $XY$, inducing a Berezinskii-Kosterlitz-Thouless ({\em BKT}) transition for perfectly isolated layers.

In our previous work, we demonstrated the principle of using molecular-based magnetism to generate new families of 2D QHAF with moderate exchange strengths. In this paper, we report the magnetic properties of a new family of molecular-based quasi-2D QHAF (Cu(pz)$_2$(ClO$_4$)$_2$, Cu(pz)$_2$(BF$_4$)$_2$, and [Cu(pz)$_2$(NO$_3$)](PF$_6$)) in which Cu$^{2+}$ ions are linked into square magnetic lattices by pyrazine molecules (pz = C$_4$H$_4$N$_2$) with exchange strengths between 10~K and 18~K. The layers are well isolated by the counter-ions, as demonstrated by their $T_{\mathrm{N}}/J$ values of 0.24 to 0.30. The saturation fields range from 300~kOe to 490~kOe so available fields are powerful enough to test the recent theoretical predictions. (The three antiferromagnets under study will sometimes be identified as the ClO$_4$, BF$_4$, and PF$_6$ compounds, respectively, in the interests of brevity.)

\begin{figure}[ht!]
\begin{center}
\epsfig{file=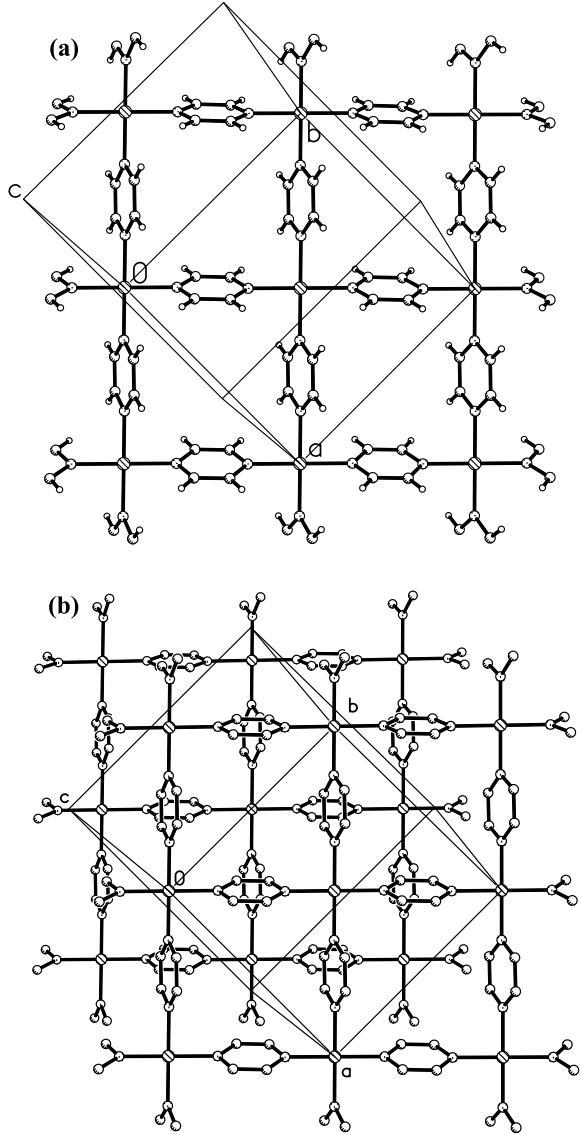,width=3in}
\caption{(a) Single layer structure of Cu(pz)$_2$(ClO$_4$)$_2$ at 293~K, viewed perpendicular to the $ab$ plane.  (b) Staggered layer structure of Cu(pz)$_2$(ClO$_4$)$_2$. The ClO$_4$ ions have been removed from the figure for clarity.}\label{layer-structure}
\end{center}
\end{figure}

\begin{figure}[ht!]
\begin{center}
\epsfig{file=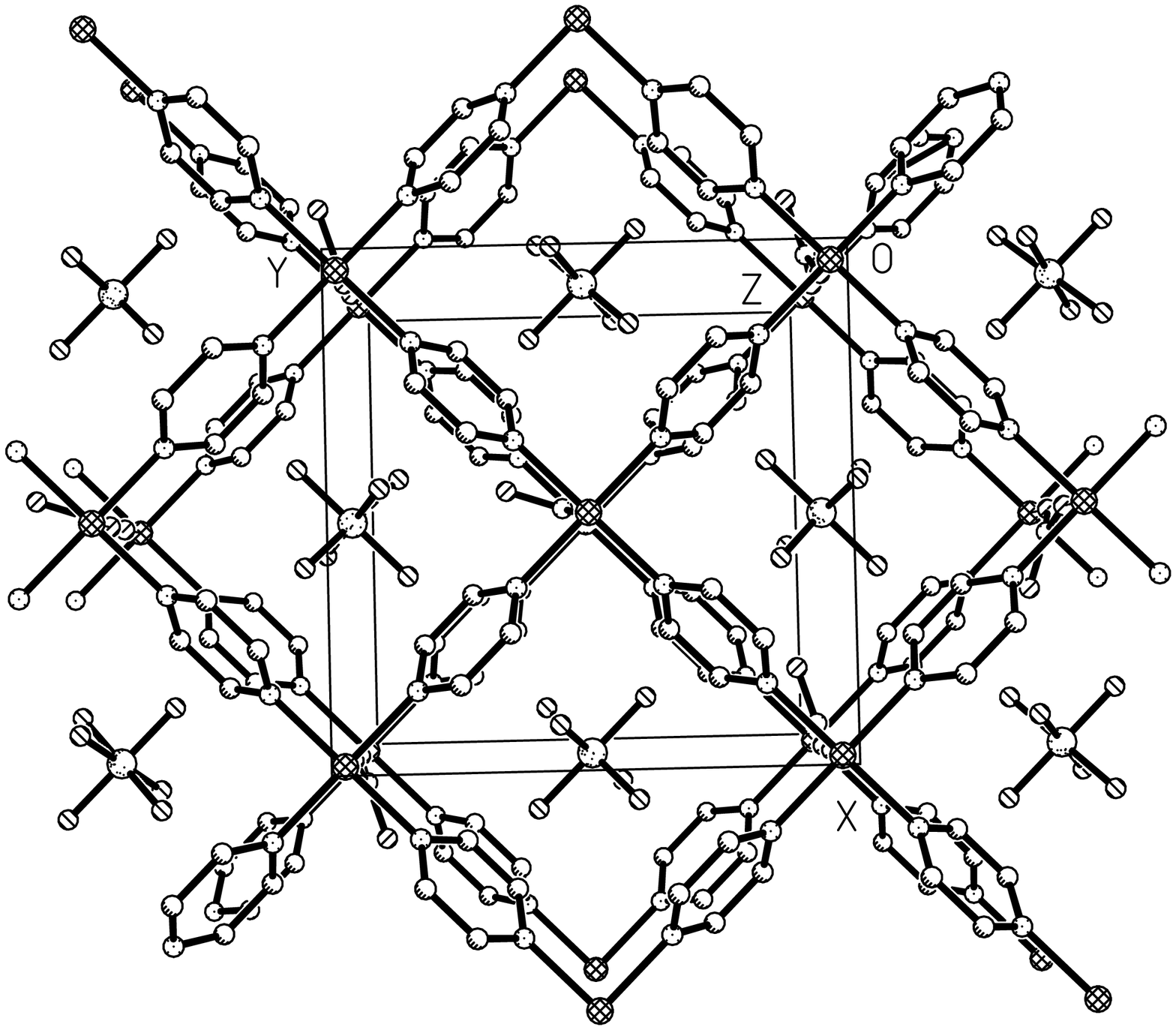,width=3in}
\caption{Staggered layer structure of [Cu(pz)$_2$(NO$_3$)](PF$_6$)}\label{tob-layer-structure}
\end{center}
\end{figure}

The three compounds consist of Cu(II) ions bonded to four neutral bridging pyrazine molecules, creating  positively charged 2D nets;  the structures are charge-balanced by the counter-ions (ClO$_4^-$, BF$_4^-$, NO$_3^-$, and  PF$_6^-$) which lie between the copper/pyrazine layers, with all anions except the PF$_6$ located in the axial sites of the copper atoms (Figs.~\ref{layer-structure} and \ref{tob-layer-structure}). A full description of the structures is found in Ref.~\onlinecite{Woodward07}.

Each of three compounds forms {\it magnetically} square lattices, even though they have three different low-temperature space groups.  Cu(pz)$_2$(ClO$_4$)$_2$ and Cu(pz)$_2$(BF$_4$)$_2$ have $C$-centered monoclinic structures at low-temperatures ($C2/c$ and $C2/m$, respectively), in which the copper sites are related by symmetry operations ($c$-glide in $C2/c$ and $C$-centering in $C2/m$) that render every copper-copper bridge equivalent even though the crystallographic angles are 96.46$^\circ$ and 120.92$^\circ$, respectively. The third compound, [Cu(pz)$_2$(NO$_3$)](PF$_6$), has a tetragonal space group $I4/mcm$ in which the coppers sit on the four-fold rotation axis, generating square 2D copper/pyrazine nets.

The copper/pyrazine layers are packed into three-dimensional lattices in two different ways, with important differences on the ultimate  low-temperature 3D ordering transitions. For both Cu(pz)$_2$(ClO$_4$)$_2$ and Cu(pz)$_2$(BF$_4$)$_2$, the counter-ions are weakly coordinated to the axial sites of the copper atoms but do not bridge between the layers. To minimize steric hindrance, adjacent layers are offset by half a unit cell in along both axes within the layer so the counter-ions can interpenetrate, Fig.~\ref{layer-structure}(b). Each copper atom is then equidistant from four coppers of the adjacent layer, an arrangement that leads to a cancellation of interlayer exchange interactions $J^\prime$.

The axial copper sites in the third compound, [Cu(pz)$_2$(NO$_3$)](PF$_6$), are occupied by bridging nitrate groups (Fig.~\ref{tob-layer-structure}) so there is a superexchange pathway between coppers of adjacent layers. For this reason, and the absence of any symmetry required cancellation of $J^\prime$, the interlayer coupling for [Cu(pz)$_2$(NO$_3$)](PF$_6$) should be the strongest of the three compounds.

\section{Experimental}
Crystals of Cu(pz)$_{2}X_2$ were grown from aqueous solutions containing a 1:2
molar ratio of Cu(ClO$_4$)$_2$ or Cu(BF$_4$)$_2$ and pyrazine.  A drop of dilute
HClO$_4$(aq) or HBF$_4$(aq) was added to the solutions to prevent precipitation of
Cu(OH)$_2$.  Crystals of $[$Cu(pz)$_2$(NO$_3$)$]$(PF$_6$) were grown from
an aqueous solution of Cu(NO$_3$)$_2$ with two equivalents of pyrazine and a 5-fold
excess of KPF$_6$ after slow evaporation over several weeks. All crystals have the forms of dark blue tablets.  Full details may be found elsewhere\cite{Woodward07}.

All DC magnetic susceptibility data were collected using Quantum Design MPMSR2 and MPMS-XL SQUID magnetometers. Susceptibility data on crystals of all three compounds were collected along all three orientation at various fields from 1.8~K to 60~K. High field magnetization data were collected on powder samples using a vibrating sample magnetometer (VSM) at the National High Field Magnet Laboratory in Tallahassee, Florida, in fields up to 30~T
at several temperatures.  Because these fields were insufficient to saturate the magnetization data were
collected in pulsed magnetic field experiments up to 60~T at the National High Field
Magnet Laboratory at LANL.  %The Electron Paramagnetic Resonance (EPR) data were
%collected on a Bruker EMX spectrometer operating at 9.4~GHz.
Zero field muon-spin relaxation (ZF $\mu^{+}$SR) measurements \cite{Blundell99,Lancaster07}
were made on powder samples of Cu(pz)$_2$(BF$_4$)$_2$ and [Cu(pz)$_2$(NO$_3$)](PF$_6$) using
the MuSR instrument at the ISIS facility, Rutherford Appleton
Laboratory, UK and using the General Purpose Surface-Muon Instrument at the Swiss Muon Source, Paul Scherrer Institute, CH. For these measurements samples were packed in Ag foil (thickness 25~$\mu$m) and mounted on a Ag backing plate.

\section{Results}
\subsection{Powder susceptibility}
\begin{figure}[ht!]
\begin{center}
\epsfig{file=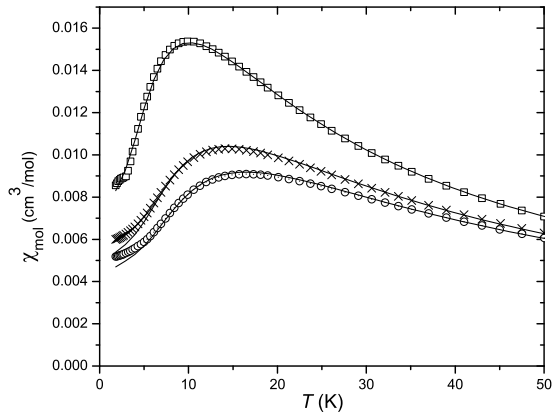,width=3in}
\caption{Powder susceptibility measured in a 1~kOe field for Cu(pz)$_2$(ClO$_4$)$_2$ ($\circ$), Cu(pz)$_2$(BF$_4$)$_2$ ($\times$) and $[$Cu(pz)$_2$(NO$_3$)$]$(PF$_6$) ($\square$). Solid lines represent fits to the 2D QHAF model.}\label{clo4-chi}
\end{center}
\end{figure}

The molar magnetic susceptibilities ($\chi_m$) as a function of temperature for powder
samples of Cu(pz)$_2$(ClO$_4$)$_2$, Cu(pz)$_2$(BF$_4$)$_2$, and $[$Cu(pz)$_2$(NO$_3$)$]$(PF$_6$) are shown in
Fig.~\ref{clo4-chi}. These data were measured in a field of 1~kOe. For each compound, a broad rounded maximum is
observed with the maximum value in $\chi_m$ occurring
near 15.9, 14.0, and 9.3 K for the ClO$_4$, BF$_4$, and $[$Cu(pz)$_2$(NO$_3$)$]$(PF$_6$) compounds respectively. The data were compared to a numerical expression for the temperature dependent susceptibility of a 2D QHAF, as determined from high temperature series expansions\cite{Navarro90} and quantum Monte Carlo
simulations\cite{troyer} to obtain exchange strengths and Curie constants as described in
Ref~\onlinecite{Woodward07}.

The solid lines are the results of the best fits of the model to the data with fitting parameters $(J,C)$ of (17.5(3)~K, 0.426(6)~cm$^{3}\cdot$K/mol), (15.3(3)~K, 0.426(6)~cm$^3\cdot$K/mol), and  (10.8(3)~K, 0.439(6)~cm$^3\cdot$K/mol) for the three compounds respectively. The quality of the fits is excellent except at the lowest temperatures where the data rise above the predicted susceptibilities. These discrepancies are discussed in detail below. The value of the Curie constant for the perchlorate compound corresponds to an average $g$-factor of 2.13, in excellent agreement with the room temperature value found from the EPR measurements. Fuller discussion of the powder susceptibilities and the magnetostructural correlations in the copper pyrazine family are found elsewhere\cite{Woodward07}.

\subsection{High field magnetization}
\begin{figure}[ht!]
\begin{center}
\epsfig{file=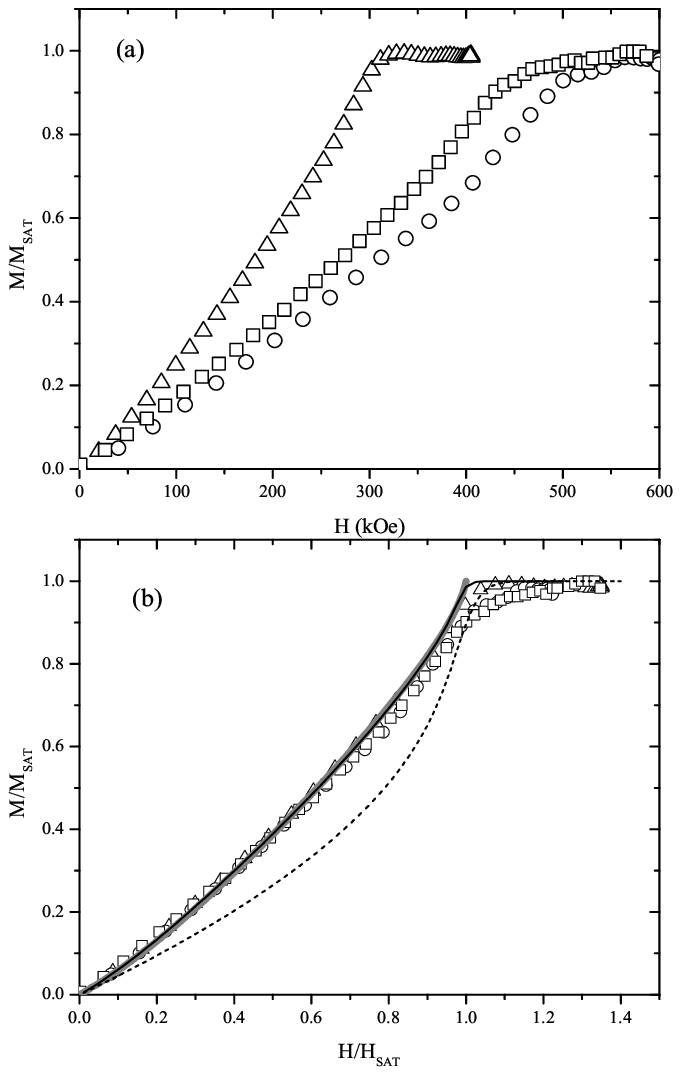,width=3in}
\caption{(a) $M/M_{\mathrm{SAT}}$ vs.\ $\mu_0 H$ at 0.50~K for Cu(pz)$_2$(ClO$_4$)$_2$ ($\bigcirc$), Cu(pz)$_2$(BF$_4$)$_2$ ($\square$) and $[$Cu(pz)$_2$(NO$_3$)$]$(PF$_6$) ($\triangle$).\\(b)
$M/M_\mathrm{SAT}$ vs.\ $H/H_\mathrm{SAT}$ at 0.50~K for Cu(pz)$_2$(ClO$_4$)$_2$ ($\bigcirc$), Cu(pz)$_2$(BF$_4$)$_2$ ($\square$), and $[$Cu(pz)$_2$(NO$_3$)](PF$_6$) ($\triangle$). The short dash line is the  $T=0$ 1D QHAF model\cite{Griffiths64}, the solid black line is a Monte Carlo Calculation
 of a 2D QHAF at $T/J = 0.05$(Ref \onlinecite{troyer2}), and the solid gray curve is a $T=0$ spinwave calculation
\cite{Zhitomirsky98}.}\label{pz-mag}
\end{center}
\end{figure}

Fig.~\ref{pz-mag}(a) shows the relative magnetization $M(H)/M_\mathrm{SAT}$ as a function of field for polycrystalline samples of all three copper pyrazine compounds. Data were collected for each compound at several temperatures between $T$=0.50~K and 4.00~K with representative points of the the $T$=0.5~K data shown in Figs.~\ref{pz-mag}(a) and \ref{pz-mag}(b). For every compound the behavior is similar: a small region in which the magnetization is linear in field, followed by a gradual upward curvature that continues to nearly the saturation value. Rounding due to thermal excitations and the variation in $g$-factor values is negligible at 0.5~K for all three samples as seen in the PF$_6$ data set. The observed rounding for the BF$_4$ and ClO$_4$ data near their saturation fields is due to their reaching $M_\mathrm{SAT}$ near the peak field when $\mathrm{d}H/\mathrm{d}t$ is small, a feature in pulsed field magnetization. Values for the saturation fields were estimated to be 310(15)~kOe, 430(20)~kOe, and 520(20)~kOe for the PF$_6$, BF$_4$, and ClO$_4$ compounds respectively, taking into consideration the rounding of the  BF$_4$, and ClO$_4$ data sets curves near saturation.
The zero-temperature saturation field of a 2D QHAF with exchange strength $J$ and four nearest neighbors can be calculated using Eq.~(\ref{hsat}). Based on the values of the exchange strengths determined by the powder susceptibility measurements and using an average g-value of 2.13 for each compound, the predicted saturation fields for the three compounds are 302(8), 428(8), and 489(8)~kOe, respectively, with the uncertainty of the calculated saturation fields due to the uncertainty of the experimental exchange constants. Within these uncertainties, the predicted $T=0$ saturation fields are equal to the experimental values  obtained at 0.5 K for the PF$_6$ and BF$_4$ compounds,  but about 6 percent lower than the experimental value of the ClO$_4$ data set. Given the lower quality of the high-field data for the ClO$_4$ compound, the value of $H_\mathrm{SAT}$ = 490~kOe obtained from Eq.~(\ref{hsat}) will be used henceforth while values of 300(10) kOe and 430(10) kOe will be used for the PF$_6$ and BF$_4$ compounds. In order to compare the three data sets to each other, the relative magnetizations were plotted as a function of the relative fields $H/H_\mathrm{SAT}$ in Fig.~\ref{pz-mag}(b). As seen in the figure, near universal behavior occurs.

Included in Fig.~\ref{pz-mag}(b) are results from numerical calculations of $M(H)(T=0)$ based on spin wave calculations for a 2D QHAF\cite{Zhitomirsky98}as a solid gray line, and a Monte Carlo simulation\cite{troyer2} of a 2D QHAF finite size lattice at $T/J = 0.05$ (solid black line). The two lines are indistinguishable except just below the saturation field. Also included is a short dash line representing the $T=0$ magnetization of a one-dimensional QHAF. See the discussion in section \ref{discussion} for details.

\subsection{Single Crystal low field Magnetization and Susceptibility}

{\bf Cu(pz)$_2$(ClO$_4$)$_2$}. Crystals of Cu(pz)$_2$(ClO$_4$)$_2$ grow as flat plates with the copper/pyrazine layers lying in the $bc$-planes, parallel to the dominant face\cite{Woodward07}. The axis normal to the layers is $a^*$ which will be defined as the $z$-direction. The two magnetically equivalent  axes in the layer are defined as $x$. The low field magnetization of single crystal of the perchlorate salt was studied at low temperatures along three orthogonal directions, two in the plane of the layers and one normal to the layers. The response of the two directions in the layers were identical, even through the crystal is monoclinic ($C2/c$ at low temperatures) and not tetragonal. The behavior normal to the layers is different.

\begin{figure}[ht!]
\begin{center}
\epsfig{file=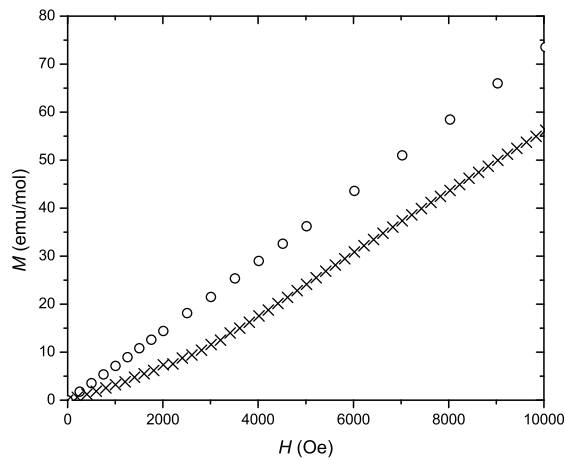,width=3in}
\caption{Low field magnetization data of Cu(pz)$_2$(ClO$_4$)$_2$ for $H \parallel x$ ($\times$) and
$H \parallel z$ ($\circ)$ at $T=1.8$~K.}\label{clo4-MvsH}
\end{center}
\end{figure}

Fig.~(\ref{clo4-MvsH}) shows the molar magnetization of Cu(pz)$_2$(ClO$_4$)$_2$ at 1.8~K between zero and 10~kOe with the fields aligned in the plane ($\times$) and normal to the plane ($\circ$). When the field is perpendicular to the layers, the magnetization increases linearly throughout the displayed range. Only for fields exceeding several tesla does the slope gradually begin to increase and display the high field behavior seen in Fig.~\ref{pz-mag}(a). In contrast, for fields within the layers, the magnetization shows two linear regions with the smaller slope occurring between 0 and 2.6(5)~kOe. No further breaks of slope occur at higher fields, only a gradual increase before saturation.

\begin{figure}[ht!]
\begin{center}
\epsfig{file=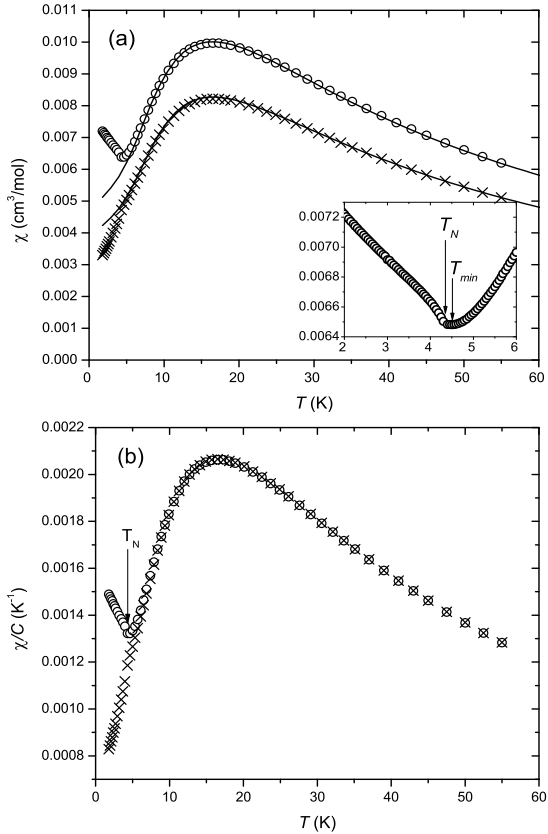,width=3in}
\caption{(a) Susceptibility of Cu(pz)$_2$(ClO$_4$)$_2$ at 1~kOe for both $H \parallel x$ ($\times)$ and $H \parallel z$ $(\circ)$. Below 5~K $\chi_z$ rises above and $\chi_x$ falls below the solid lines representing the susceptibility for an ideal 2D QHAF. The inset shows the details of $\chi_z$ between 2~K and 6~K. (b) Susceptibility data normalized by Curie constant for Cu(pz)$_2$(ClO$_4$)$_2$, same notation as in (a).}\label{clo4-chi-both}
\end{center}
\end{figure}

The two susceptibilities, $\chi_x$ and $\chi_z$, as determined in a field of 1~kOe, are shown at low temperatures in Fig.~\ref{clo4-chi-both}. Two types of anisotropy are observed, with the first being the simple $g$-factor anisotropy $g_z = 2.27$ and $g_{x} = 2.07$ as determined by EPR. The corrected susceptibilities,  $\chi_i/C_i$, are shown in the lower panel of the figure where the second, temperature dependent, anisotropy is evident. As the temperature drops below that of the susceptibility maximum, $\chi_x$ decreases steadily, though not to zero, while a minimum is found in $\chi_z$ at 4.5~K; for temperatures below the minimum, the susceptibility increases linearly upward. Zero-field AC-susceptibility measurements confirm the 1~kOe DC susceptibility presented in Fig~\ref{clo4-chi-both}. As will be discussed below, the minimum is neither the signature of long-range order nor a field-induced effect, but is due to an intrinsic spin-anisotropy  crossover.

{\bf Cu(pz)$_2$(BF$_4$)$_2$}. The behavior of Cu(pz)$_2$(BF$_4$)$_2$ at low fields is very similar to Cu(pz)$_2$(ClO$_4$)$_2$, with the magnetization curve showing a slope change at around 2.5(6)~kOe for fields  within the layer while no such breaks were observed for a field perpendicular to the layer. The $M/H$ vs.\ $T$ data also show a minimum around 4.05~K only when the field is parallel to the layer.

\begin{figure}[ht!]
\begin{center}
\epsfig{file=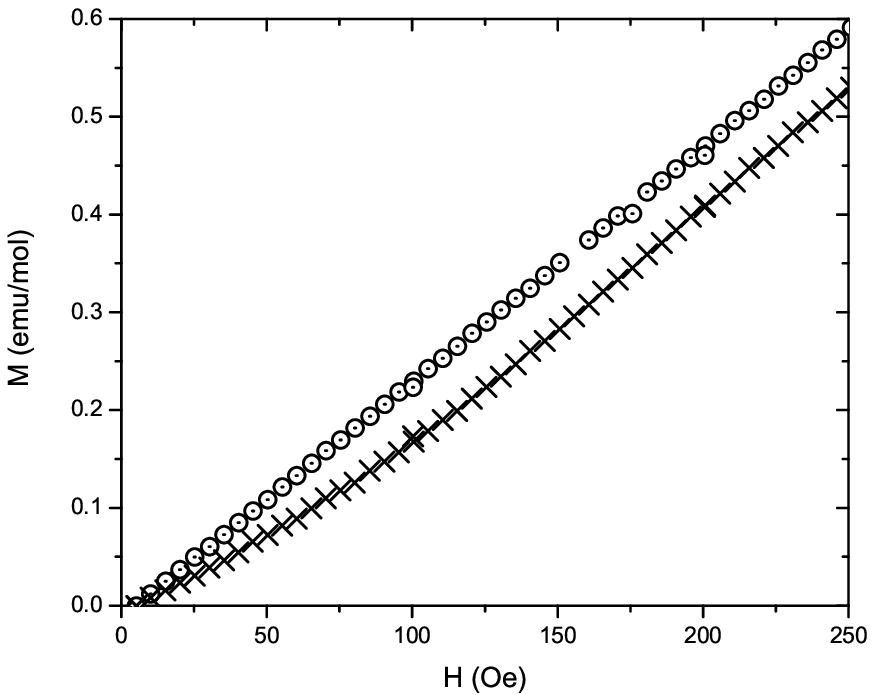,width=3in}
\caption{Single crystal magnetization data (T = 1.8 K) of $[$Cu(pz)$_2$(NO$_3$)$]$(PF$_6$) crystal for $H \parallel x$ ($\times$) and $H \parallel z$ ($\circ$).}\label{tob-MvsH}
\end{center}
\end{figure}

{\bf [Cu(pz)$_2$(NO$_3$)$]$(PF$_6$)}. This tetragonal compound also grows as flat plates. The normal to the plate is the $c$-axis and will be referred to as the $z$-direction. The equivalent axes in the plane are the tetragonal $a$-axes and point in the $x$-direction. The magnetization as a function of field of the PF$_6$ compound at 1.8~K is shown in Fig.~\ref{tob-MvsH}. A field-induced transition at around 70~Oe is observed for a field along $x$ axis takes place while no such anomaly occurs for fields in the $z$ direction. This behavior is similar to that seen for the ClO$_4$ and BF$_4$ compounds but at a substantially smaller field. For fields greater than 100~Oe the magnetization along each axis is linear up to 5~kOe, then gradually curves upward. $M_z$ is greater than $M_x$ for the same field and temperature, consistent with the greater value of $g_z$.

\begin{figure}[ht!]
\begin{center}
\epsfig{file=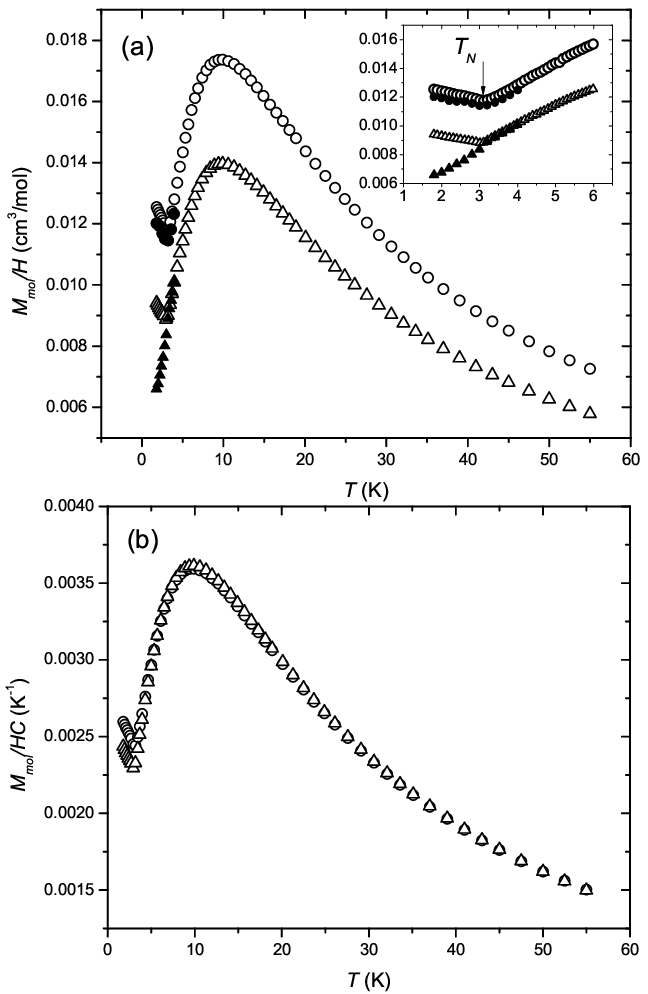,width=3in}
\caption{(a) $M/H$ ratios of [Cu(pz)$_2$(NO$_3$)$]$(PF$_6$) for both $H \parallel x$ ($\triangle)$ and $H \parallel z$ ($\circ)$. The inset shows details of $M/H$ for both directions between 2~K and 6~K. Open symbols represent DC data at 1~kOe and filled symbols represent AC data measured at zero-field. 
(b) DC $M/H$ data normalized by Curie constant for [Cu(pz)$_2$(NO$_3$)$]$(PF$_6$), same notation as in (a).}\label{tob-chi-both}
\end{center}
\end{figure}

The zero-field ac susceptibilities for [Cu(pz)$_2$(NO$_3$)$]$(PF$_6$) (filled symbols in Fig.~\ref{tob-chi-both}(a) have similar temperature dependences as those of both the the zero-field ac susceptibilities and DC susceptibilities of Cu(pz)$_2$(ClO$_4$)$_2$ (Fig~\ref{clo4-chi-both}); there is a minimum for $\chi_z$ but no minimum in $\chi_x$. However, for [Cu(pz)$_2$(NO$_3$)$]$(PF$_6$) in DC fields of 100~Oe or greater, a minimum appears for $\chi_x$ as well. Data measured in a field of 1~kOe are shown in Fig.~\ref{tob-chi-both}. The mimimum for $\chi_z$ is intrinsic and due to an internal anisotropy, as is the case for Cu(pz)$_2$(ClO$_4$)$_2$; the addition of the DC-field modifies the minimum but does not create it. In contrast, the  minimum for $\chi_x$ in the 1~kOe field is not intrinsic, but is field-induced. An analysis of the two types of anisotropies is found in the Discussion.

\subsection{Determination of $T_{\mathrm{N}}$}
The unambiguous identification of long range magnetic order in low-dimensional systems is often
made difficult by the existence of quantum fluctuations, which act to depress the ordering temperature
and reduce the size of the magnetic moments. This often hinders experimental methods such as
magnetic susceptibility or magnetic neutron diffraction.
In addition, the build up of spin correlations above the ordering temperature in these systems
reduces the available entropy and hence the response of heat capacity\cite{Lancaster07}. In contrast,
muon-spin relaxation measurements have been shown to detect magnetic order in cases where transitions
are very difficult to observe with more conventional techniques\cite{Blundell07}.

{\bf Cu(pz)$_2$(ClO$_4$)$_2$} The zero-field N\'{e}el temperature of Cu(pz)$_2$(ClO$_4$)$_2$ has recently been determined to be 4.21(1) K using $\mu^{+}$SR \cite{Lancaster07}. Using  single crystal magnetometry, we have confirmed this value for the ClO$_4$ compounds and have also precisely ascertained the ordering temperatures for the BF$_4$ and PF$_6$ compounds.

In Fig.~\ref{clo4-chi-both}, starting at 1.8~K, for 1~kOe, the susceptibility along $z$-direction drops as the temperature rises until, at a temperature just below that of the minimum, the slope of the curve drops sharply at 4.25 K, before undergoing a change of curvature with the slope growing more positive (see inset, Fig.~\ref{clo4-chi-both}(a)). The sudden change of curvature occurs at the same temperature at which the muon experiment \cite{Lancaster07} found the internal fields of Cu(pz)$_2$(ClO$_4$)$_2$ to vanish, within experimental error.

\begin{figure}[ht!]
\begin{center}
\epsfig{file=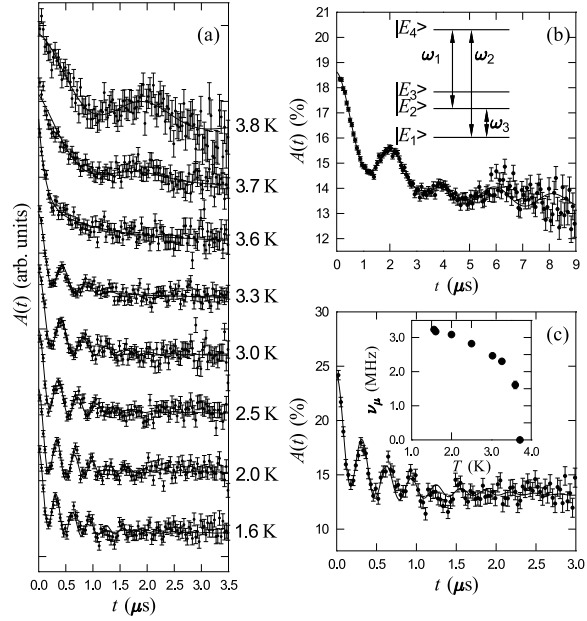,width=3in}
\caption{(a) Temperature evolution of ZF $\mu^{+}$SR spectra measured
on Cu(pz)$_2$BF$_4$. (b) Above
$T_{\mathrm{N}}$ low frequency oscillations are observed due to the
dipole-dipole coupling of F--$\mu^{+}$--F states. {\it Inset:}
The energy level structure allows three transitions, leading to
three observed frequencies. (c) Below $T_{\mathrm{N}}$
higher frequency oscillations are observed
due to quasistatic magnetic fields at the muon sites.
{\it Inset:} The evolution of the muon precession frequency $\nu_{\mu}$ with temperature.
 }\label{bf4-muon}
\end{center}
\end{figure}

{\bf Cu(pz)$_2$(BF$_4$)$_2$}. ZF $\mu^{+}$SR spectra measured on Cu(pz)$_2$(BF$_4$)$_2$ at
several temperatures are shown in Fig.~\ref{bf4-muon}(a).
Below $T_{\mathrm{N}}$ (Fig.~\ref{bf4-muon}(c))
we observe oscillations in the time dependence of the muon
polarization (the ``asymmetry'' $A(t)$ ) which are
characteristic of a quasi-static local magnetic field at the
muon stopping site. This local field causes a coherent precession of the
spins of those muons for which a component of their spin polarization
lies perpendicular to this local field (expected to be 2/3 of the
total spin polarization for a powder sample).
The frequency of the oscillations is given by
$\nu_{i} = \gamma_{\mu} |B_{i}|/2 \pi$, where $\gamma_{\mu}$ is the muon
gyromagnetic ratio ($=2 \pi \times 135.5$~MHz T$^{-1}$) and $B_{i}$
is the average magnitude of the local magnetic field at the $i$th muon
site. The precession frequency $\nu$ is proportional to the order
parameter of the system.
Any fluctuation in magnitude of these local fields will
result in a relaxation of the oscillating signal, described by
relaxation rates $\lambda_{i}$.
The
presence of these oscillations at low temperatures in Cu(pz)$_2$(BF$_4$)$_2$ suggests
very strongly that this material is magnetically ordered below $T_{\mathrm{N}}=3.7(1)$~K.
Oscillations are observed at a single frequency, suggesting a single muon
site in this material (in contrast to the case of
the ClO$_{4}$ material, where three frequencies were observed\cite{Lancaster07}). We note,
however,
 that the oscillations are quite heavily damped (with a typical relaxation rate
of $\lambda \approx 1.7$~MHz) compared with those
measured in the  ClO$_{4}$ case (where $\lambda \approx 0.4$~MHz) which may imply the
presence of several muon sites experiencing slightly different local magnetic
fields.

Above $T_{\mathrm{N}}$ the character of the measured spectra changes
considerably (Fig.\ref{bf4-muon}(a) and (b)) and we observe lower
frequency oscillations
characteristic of the dipole-dipole interaction of the muon
and the $^{19}$F nuclei comprising the BF$_{4}$ counter-ions. This
behavior\cite{Brewer86} has been observed previously in systems of this kind\cite{Lancaster07prl}.
Fits to the data show that the muon stops between
two fluorine atoms and
the resulting
 F--$\mu^{+}$--F spin system consists of four distinct
energy levels with three allowed transitions between them
(inset, Fig.~\ref{bf4-muon}(b))
giving rise to the distinctive three-frequency oscillations observed. Fits of the data
suggest a $\mu^{+}$-F separation of 0.12~nm and an F--$\mu^{+}$--F bond angle of $\sim 140^{\circ}$.

Variable temperature measurements  of the magnetization of single crystals of Cu(pz)$_2$(BF$_4$)$_2$ (not shown) have revealed behavior similar to that seen in the ClO$_4$ compound shown in  Fig.~\ref{clo4-chi-both}, although at slightly lower temperatures.  A slope anomaly indicative of the ordering temperature and the temperature of $\chi_\mathrm{min}$ of Cu(pz)$_2$(BF$_4$)$_2$ have been found to be 3.80(5)~K and 4.05(5)~K, respectively.

{\bf [Cu(pz)$_2$(NO$_3$)$]$(PF$_6$)}. Variable temperature studies of single crystals of the PF$_6$ compound (inset of Fig.~\ref{tob-chi-both}) show that the ordering temperature and the temperature of $\chi_\mathrm{min}$ are very close to each other, both occurring at 3.05~K.
The ordering temperature of [Cu(pz)$_2$(NO$_3$)$]$(PF$_6$), as determined by $\mu^+$SR experiments, has been previously reported\cite{Lancaster07prl} to be near 2.0(2)~K.
The significant disagreement between the two values for the critical temperature of [Cu(pz)$_2$(NO$_3$)$]$(PF$_6$) is in sharp contrast to the excellent agreement found from the two techniques for Cu(pz)$_2$(ClO$_4$)$_2$ and the good agreement found for Cu(pz)$_2$(BF$_4$)$_2$.
It is worth noting, however, that the oscillations observed
for this material due the the magnetic ordering
were very heavily damped (with $\lambda \approx 4$~MHz),
making the unambiguous determination of an ordering temperature
very difficult in this case. The previous estimate obtained from the muon measurements, corresponding to the
disappearance of the magnetic oscillations (due to the relaxation rate increases with increasing temperature),
should therefore be taken as
a lower bound on $T_{\mathrm{N}}$. The heavy damping observed in this material is suggestive of a significantly
larger  width of the local magnetic field distribution in [Cu(pz)$_2$(NO$_3$)$]$(PF$_6$)  compared to
the ClO$_{4}$ and PF$_{6}$ materials. Our $\mu^{+}$SR measurements do, however, unambiguously confirm the presence of long
range magnetic order in this material at low temperature.
For the purposes of this paper, we will consider the N\'{e}el temperatures to correspond to the temperatures of the slope anomalies in $\chi_z$.

\section{Discussion}\label{discussion}
\subsection{High Field Magnetism}
The $T=0.5$~K magnetizations of ClO$_4$, BF$_4$, and PF$_{6}$ materials are qualitatively similar; They begin with a small initial slope that gradually increases with field until the maximum slope is reached just before saturation; the respective saturation fields as based on the exchange constants of the three compounds are calculated to be 490, 430, and 300~kOe (Fig.~\ref{pz-mag}(a)). When appropriately normalized as $H/H_\mathrm{SAT}$, the three compounds show universal behavior (Fig.~\ref{pz-mag}(b)). Similar behavior was observed in a previous study\cite{Woodward01, Woodward02} of several less well-isolated copper bromide 2D QHAF.

The $T = 0$ magnetization for the 2D QHAF has been calculated using a spin-wave expansion\cite{Zhitomirsky98} and is displayed as the solid gray curve in Fig.~\ref{pz-mag}(b). It is qualitatively similar to the normalized data but has a somewhat smaller slope for the first half of the magnetization curve and rises to an infinite slope at the saturation field. In addition, the magnetization curve at a relative temperature $k_{\mathrm{B}}T/J = 0.05$ has been calculated using a quantum Monte Carlo simulation\cite{troyer} and plotted as the solid black line in Fig.~\ref{pz-mag}(b). (This simulated result is identical to the $T = 0$ calculation except for values of $H/H_\mathrm{SAT}>0.9$ where is slightly reduced due to the thermal excitation of spin waves.) The discrepancy between the experimental and theoretical slopes is attributed to the fact that all three compounds are in the 3D ordered state at $T = 0.5~K$ while the model describes an ideal 2D QHAF. As seen in Fig.~\ref{clo4-chi}, the experimental low-temperatures susceptibilities are all higher than those predicted by the 2D QHAF model. The magnetization data of these 2D compounds display much less curvature that appears in the 1D QHAF, represented by the short dash line in Fig.~\ref{pz-mag}(b); the difference arises from the greater influence of quantum fluctuations in the 1D system which reduces the effective moments.

High-field magnetization of a similar series of copper/pyrazine 2D antiferromagnets ([Cu(pz)$_2$(HF$_2$)]($X$), where $X =$ClO$_4$, BF$_4$, PF$_6$, SbF$_6$) have recently\cite{Goddard08} been reported on both single crystal and polycrystalline samples. These results for the polycrystalline samples are similar to those displayed in Fig.~\ref{pz-mag}  but the magnetization curves for single crystals are much less rounded near the saturation field, a consequence of the presence of a single $g$-factor. For typical copper compounds, the $g$-factors range from 2.07 to 2.27, leading to a spread of critical fields of roughly ten percent in studies of randomly oriented samples.

\subsection{Interlayer Exchange and Intrinsic Anisotropies}
\begin{table*}[t!]
   \caption{Parameters for
the  layered compounds Cu(pz)$_2$(ClO$_4$)$_2$, Cu(pz)$_2$(BF$_4$)$_2$,
[Cu(pz)$_2$(NO$_3$)](PF$_6$) and Sr$_2$CuO$_2$Cl$_2$.\label{comparison-table}}
     \begin{ruledtabular}
   \begin{tabular}{llllllllll}
&  $J$ & $T_{\mathrm{N}}$ &$k_{\mathrm{B}} T_{\mathrm{N}}/J$ & $J^\prime/J$ \footnote{Based on Eq.~(\ref{Yasuda}) which assumes $\Delta=0$} &$H_{\mathrm{A}}$ &$H_{\mathrm{SAT}}$
&$H_\mathrm{A}/H_{\mathrm{SAT}}$ &$\Delta_\mathrm{CO}$ \\
&  (K) & (K) & &   &(kOe) &(kOe)&  &\\
\hline
Cu(pz)$_2$(ClO$_4$)$_2$& 17.5 & 4.25 & 0.243 & 8.8 $\times 10^{-4}$ &2.6 &490 & 5.3$\times$10$^{-3}$ &4.6$\times$10$^{-3}$\\
Cu(pz)$_2$(BF$_4$)$_2$ &15.3 & 3.8 & 0.248  &1.1$\times 10^{-3}$ &2.5 &430 & 5.8$\times$10$^{-3}$ & 6.2$\times$10$^{-3}$\\
$[$Cu(pz)$_2$(NO$_3$)$]$(PF$_6$) & 10.8 &3.05 &0.282 & 3.3$\times 10^{-3}$ & 0.07 & 300 & 2.3$\times$10$^{-4}$ &1.2$\times$10$^{-2}$\\
Sr$_2$CuO$_2$Cl$_2$ &1450  &251 &0.173  &1.9$\times 10^{-5}$ &7 &40000 &1.8$\times$10$^{-4}$ &3.6$\times$10$^{-4}$\\
   \end{tabular}
     \end{ruledtabular}
\end{table*}

A figure of merit of characterizing low-dimensional antiferromagnets is the critical ratio of the N\'{e}el temperature to the dominant exchange strength, $T_{\mathrm{N}}/J$. For Cu(pz)$_2$(ClO$_4$)$_2$, Cu(pz)$_2$(BF$_4$)$_2$, and [Cu(pz)$_2$(NO$_3$)](PF$_6$), the critical ratios are 0.24, 0.25 and 0.28, respectively (Table~\ref{comparison-table}). The ideal 2D QHAF has a critical ratio of zero\cite{Manousakis89} so the ordering present in these compounds must arise from lattice anisotropy (3D interaction $J^\prime$), exchange anisotropy ($\Delta\neq0$ in the exchange Hamiltonian in Eq.~(\ref{Hamiltonian})) or both. As will be shown below, it is not possible to determine an accurate value of $J^\prime$ in the presence of a nonzero $\Delta$.

An alternative measure of the degree of isolation for {\it quasi}-2D QHAF is given by the magnetic correlation length $\xi$ at the ordering temperature  $\xi(T_{\mathrm{N}})$). The correlation length  is known to diverge exponentially at low temperature with only a weak temperature dependence in the prefactor\cite{Chakravarty88}. The full expression for the correlation length in units of the lattice constant $a$ is given by\cite{Hasenfratz91}
\begin{equation}
\frac{\xi}{a}= \frac{e}{8} \frac{c/a}{2\pi\rho_{s}}\exp \left( \frac{2\pi\rho_{s}}{T} \right)
\left( 1-0.5\frac{T}{2\pi\rho_{s}}+O(\frac{T}{2\pi\rho_{s}})^{2} \right)
\label{cor}
\end{equation}
where $c=1.657Ja$ and $\rho_s=0.1830 J$  are the renormalized spin wave velocity and spin-stiffness constants, respectively.  For Cu(pz)$_2$(ClO$_4$)$_2$ with a critical ratio of 0.24, the correlation length at the ordering temperature is predicted to be  $\xi(T_{\mathrm{N}})/a = 50$, while the values for Cu(pz)$_2$(BF$_4$)$_2$, and [Cu(pz)$_2$(NO$_3$)](PF$_6$) are calculated to be 45 and 25, respectively. For comparison, the correlation lengths $\xi/a$ at $T_{\mathrm{N}}$ for Sr$_{2}$CuO$_{2}$Cl$ _{2}$ and deuterated copper formate tetrahydrate have been determined by neutron scattering experiments to
be approximately 220 (Ref.~\onlinecite{Greven95}) and 55 (Ref.~\onlinecite{Ronnow99}), respectively.

The ideal 2D QHAF only orders\cite{Manousakis91} at $T = 0$, but all {\it quasi}-2D QHAF contain finite interlayer couplings $J^\prime$ that induce long-range order (LRO) at a finite temperature $T_{\mathrm{N}}$.  Recently a method for estimating the interlayer coupling constant $J^\prime$ in 3D arrays of isotropic 2D QHAF has been developed based on a modified random phase approximation, modeled with classical and quantum Monte Carlo simulations\cite{Yasuda2005}. The approach leads to an empirical formula relating $J^\prime$ and $T_{\mathrm{N}}$,

\begin{equation}
\frac{J^\prime}{J}=\exp \left( b-\frac{4\pi\rho_s}{T_{\mathrm{N}}} \right),
\label{Yasuda}
\end{equation}
where $\rho_s$ is the spin stiffness ($\rho_s=0.183J$ for the 2D QHAF\cite{Beard98}) and $b = 2.43$ for $S=1/2$. This result shows that $T_{\mathrm{N}}/J$ decreases only logarithmically with $J^\prime/J$ ratio; even very well isolated 2D layers will have critical ratios far from zero. For $J^\prime/J$ ratios of 10$^{-1}$, 10$^{-2}$, 10$^{-3}$, the corresponding values of $T_{\mathrm{N}}/J$ are 0.491, 0.326, and 0.244. The 2D QHAF known to have the greatest degree of isolation is Sr$_2$CuO$_2$Cl$_2$ with a $T_{\mathrm{N}}/J$ = 256.5~K/1450~K = 0.18 (Ref.~\onlinecite{Greven95}), corresponding to $J^\prime/J$=3$\times$10$^{-5}$. For La$_2$CuO$_4$, $T_{\mathrm{N}}/J$ = 311~K/1500~K = 0.207, corresponding to $J^\prime/J$= 1.7$\times$10$^{-4}$.  Using Eq.~(\ref{cuccolieqn}) and the $T_\mathrm{N}/J$ ratios for the ClO$_4$, BF$_4$ and PF$_6$ compounds, their $J^\prime/J$ ratios are found to be 8.8$\times$10$^{-4}$, 1.1$\times$10$^{-3}$, and 3.3$\times$10$^{-3}$ respectively.

The values of $J^\prime$ obtained from Eq.~(\ref{Yasuda}) are not necessarily correct for a given compound, since they are calculated under the assumption that the exchange interactions are strictly Heisenberg, $\Delta =0$. Purely Heisenberg magnetic behavior can be found in systems of organic radicals or transition metal ions with spin-only moments, due either to half filled shells or complete quenching of orbital angular momentum. The $3d^5$ Mn(II) has a half-filled shell and a spin-only moment, as ascertained by its $g$-factor of 2.00. However, the orbital angular momentum of the $3d^9$ Cu(II) ion in a non-cubic site is not completely quenched; enough orbital angular momentum remains to create $g$-factors ranging typically from 2.05 to 2.25 for different orientations of the external field relative to the coordination geometry. These remaining internal fields at the copper site typically lead to anisotropy parameters in the range $-0.02<\Delta<0.02$.

Evidence for exchange anisotropy is found in the low-temperature, low-field magnetization curves (Figs.~\ref{clo4-MvsH} and \ref{tob-MvsH}). For all three compounds, changes of slope in the magnetization curve are found when the field is applied within the copper/pyrazine planes; no magnetization anomalies occur for fields normal to the planes. These anisotropy fields, labeled as $H_{\mathrm{A}}$, are 2.6~kOe, 2.5~kOe, and 70~Oe for the ClO$_4$, BF$_4$, and PF$_6$ compounds, respectively, Table~\ref{comparison-table}. Normalized by their respective saturation fields $H_{\mathrm{SAT}}$ of $\approx 490$~kOe, 430~kOe, and 300~ kOe, the anisotropy ratios
$H_{\mathrm{A}}/H_{\mathrm{SAT}}$, are  $5.3 \times 10^{-3}$, $5.8 \times 10^{-3}$, and $2.3 \times 10^{-4}$
(Table~\ref{comparison-table}). The ratios for ClO$_4$ and BF$_4$ are close to a percent but the value for the PF$_6$ is surprisingly more than a factor of thirty smaller. These anomalies point to an $XY$ anisotropy as they occur for fields applied within copper pyrazine planes, normal to the anisotropy axis; no anomalies are found for fields along the $z$-axes.

Additional evidence for $XY$ anisotropy is found in the single crystal susceptibilities. Were Ising anisotropy present, the easy axis susceptibility would descend to zero as the temperature drops below the N\'{e}el temperature. In the presence of an $XY$ anisotropy, all susceptibilities remain finite in the zero-temperature limit due to the continuous rotational symmetry of the ground state in the $xy$ plane; it is this behavior that is observed for all measured susceptibilities. Yet it is the minima in the susceptibilities $\chi_z$ for each compound that provide the clearest evidence of $XY$-anisotropy.

The minimum in the out-of-plane component of the uniform susceptibility of a 2D QHAF has previously been recognized as a signature of  $XY$ anisotropy \cite{Cuccoli_PRB_2003, Cuccoli_PRL_2003}. As discussed below, the temperature of the minimum in $\chi_z$, denoted as  T$_{CO}$, marks the crossover from Heisenberg to $XY$ behavior. Based on the ratio of T$_{CO}$ to the  intralayer exchange constant $J$, Cuccoli and coworkers have obtained an empirical formula \cite{Cuccoli_PRL_2003} which can be used obtain an estimate for the exchange anisotropy parameter  $\Delta_{CO}$.
\begin{equation}
T_\mathrm{CO}\approx\frac{4\pi\cdot0.214J}{\ln(160/\Delta_\mathrm{CO})}
\label{Cuccoli_co}
\end{equation}

Implementation of that formula for the  ClO$_4$, BF$_4$, and PF$_6$ compounds leads to values  $\Delta_{CO}$ which are $4.6 \times$10$^{-3}$, $6.2 \times 10 ^{-3}$, and $1.2 \times 10 ^{-2}$ (Table~\ref{comparison-table}). We note that these values for the anisotropy parameters for the ClO$_4$ and BF$_4$ compounds are very close to the values of the respective field ratios $H_{\mathrm{A}}/H_{\mathrm{SAT}}$ and conclude that the minimum in $\chi_z$ does provide a quantitative measure of the degree of $XY$-anisotropy in {\it quasi}-2D QHAF.

For the PF$_6$ compound the values of $H_\mathrm{A}/H_\mathrm{SAT}$ (2.3$\times$10$^{-4}$) and $\Delta_\mathrm{CO}$ (1.2$\times$10$^{-2}$) differ by a factor of 50. The compound with the smallest anisotropy has the highest relative crossover and ordering temperatures. It is likely that this discrepancy is due to a high value of $J^\prime/J$ for the PF$_6$ compound. Eq.~(\ref{Cuccoli_co})  was derived on the assumption that $J^\prime$= 0 so the existence of a minimum in $\chi_z$ would only be due to an exchange anisotropy. For the PF$_6$ compound the 3D ordering temperature dominates and prevents the material from ever reaching a low enough temperature for the exchange crossover to appear.

There is a simple qualitative explanation for the appearance of the minimum in $\chi_z$ in a magnetic system with $XY$ anisotropy. At high temperatures the system is isotropic with the orientation of antiferromagnetically coupled pairs fluctuating in all three directions. $\chi_x$ and $\chi_z$ are equal (within $g$-factor anisotropy) and they decrease equally with decreasing temperature ($T < J$) as there is less thermal energy to overcome the antiferromagnetic coupling. As the temperature is lowered further, the  $XY$ anisotropy becomes increasingly relevant and an larger fraction  of the spins anti-align in the plane. For a field in the z-direction, this process increases the fraction of antiferromagnetically coupled spins which will cant in the direction of the field; consequently, $\chi_z$ begins to increase as the temperature cools and its minimum appears. In contrast, for a field in the plane, the number of responding spin pairs canting in the direction of the field decreases (there are fewer moments along +/- z) and $\chi_x$ falls below the value of the isotropic 2D QHAF (See Fig.~\ref{clo4-chi-both}(a)). The minimum in $\chi_z$ marks the temperature at which the out-of-plane component of the antiferromagnetic coupling becomes irrelevant and the system crosses over from Heisenberg to $XY$ anisotropy.

In light of the presence of the $XY$ anisotropy, the physical significance of the anisotropy fields becomes apparent. At the low temperature of 1.8~K, the antiferromagnetically-coupled moments fluctuate within the copper/pyrazine planes. Application of a field within the plane will cant the moments oriented primarily normal to the field, but there will be little response of the moments along the axis of the field. As the field is increased towards $H_{\mathrm{A}}$ the Zeeman energy overcomes the anisotropy energy and the moments are now free to rotate out of the easy plane. This dramatically increases the number of moments capable of canting in the direction of the field; the slope of the $M$ vs.\ $H$ curve should increase substantially for fields greater than $H_{\mathrm{A}}$. Experimentally the ratios of the $M$ vs.\ $H$ slopes above the critical field to the slopes below have been found to be 1.94, 1.68, and 1.37 for the ClO$_4$, BF$_4$ and PF$_6$ compounds, respectively. The small value for the PF$_6$ compound is consistent with its much smaller anisotropy parameter.

A $\chi_z$ minimum has previously been reported\cite{Miller1990} for Sr$_2$CuO$_2$Cl$_2$, the  2D QHAF with the lowest critical ratio, (Table~\ref{comparison-table}). Magnetic susceptibility and  $^{35}$Cl nuclear spin-lattice relaxation rate studies\cite{Borsa1992PRB,Suh1995PRL} were interpreted as demonstrating the existence of a small amount of $XY$ anisotropy. According to Eq.~(\ref{Cuccoli_co}), the ratio of crossover temperature to exchange strength for this compound (300~K/1450~K = 0.21) corresponds to $\Delta_\mathrm{CO} = 3.6\times10^{-4}$. Measurements of the out-of-plane spin wave gap by neutron scattering\cite{Greven95}  directly established the value of  the anisotropy parameter to be $\Delta_\mathrm{exp} =1.4 \times 10^{-4}$, in very good agreement with the formula. Similar experiments are required to directly determine the anisotropy parameters of the copper/pyrazine antiferromagnets.

\subsection{Long range order}

In the presence of exchange anisotropy, two-dimensional magnets order at finite temperatures even in the absence of 3D interactions. If $\Delta < 0$, the $S^x$ and $S^y$ components are more heavily weighted than $S^z$; in the limit $\Delta=1$, the spin degrees of freedom have been totally reduced from three to two and spontaneous order occurs \cite{Berezinskii71, Kosterlitz73} at the Berezinskii-Kosterlitz-Thouless ({\em BKT}) transition temperature, $T_\mathrm{BKT} = 0.90JS^2$ for classical spins and $0.353J$ for $S=1/2$ (Ref \onlinecite{Ding90PRB}). This unique order is characterized by a diverging susceptibility but no spontaneous magnetization. No magnetic system is known to undergo a such a transition but BKT behavior has been observed in superfluid or superconducting films, as well as in arrays of Josephson junctions\cite{Mooij94}. Similarly, for Ising-like anisotropy ($\Delta<0$), the increased weighting of the axial components leads to spin-spin correlations diverging at finite temperatures. In the presence of complete Ising anisotropy ($\Delta\rightarrow\infty$), $T_C = 2.269J$ (Ref \onlinecite{LOnsager44}). Given the importance of exchange anisotropy in determining the critical ratio, no estimate of $J^\prime$ is possible without first determining the sign and strength of $\Delta$. We have done so for Cu(pz)$_2$(ClO$_4$)$_2$, Cu(pz)$_2$(BF$_4$)$_2$, and [Cu(pz)$_2$(NO$_3$)](PF$_6$) in the previous section.

Can the presence of a {\em small} $XY$ anisotropy ($\sim$10$^{-2}$) induce long-range order? While pure $XY$ ansiotropy ($\Delta=1$ in Eq.~(\ref{Hamiltonian})) leads to the BKT transition, quantum fluctuations have a larger influence near the Heisenberg limit of $\Delta = 0$. It was not known that small $XY$ anisotropies were capable of stablizing order until the QMC studies of Ding\cite{Ding1992PRL}. Simulations on lattices up to $96 \times 96$ sites demonstrated that BKT order is surprisingly stable in the presence of small anisotropies. For anisotropy values of 0.02, 0.1, and 0.5, the critical temperatures were found to be 0.250$J$, 0.285$J$, and 0.325$J$, respectively. Scaling arguments show that the naive guess that the critical temperatures should scale with the anisotropy parameter is incorrect; instead, $T_{\mathrm{BKT}}\sim 1/\ln(C/\Delta)$. More recent simulations on larger lattices\cite{Cuccoli_PRB_2003} confirmed the conclusions of Ding and extended the study to an anisotropy as small as $10^{-3}$, finding that $T_{\mathrm{BKT}} =0.175 J$ and 0.229$J$ for anisotropies of $10^{-3}$ and 0.02. The dependence of their critical temperatures upon the anisotropy parameter can be expressed by the second empirical equation
\begin{equation}
T_{\mathrm{BKT}}/J = \frac{2.22}{\ln(330/\Delta)}
\label{cuccolieqn}
\end{equation}

\begin{figure}[ht!]
\begin{center}
\epsfig{file=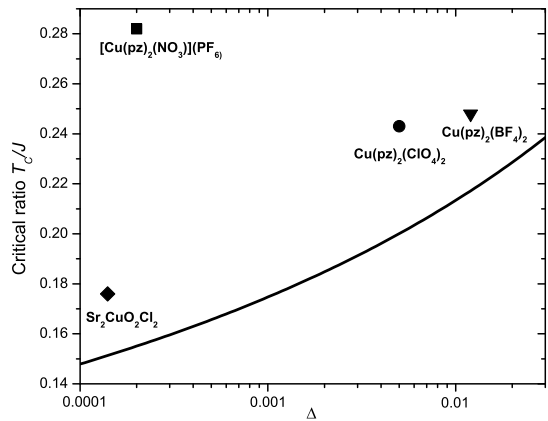,width=3in}
\caption{The critical ratio $T_{\mathrm{N}}$/J is plotted against the exchange anisotropy parameter $\Delta=0$. The solid line is the prediction of Eq.~(\ref{cuccolieqn}) which assumes $J^\prime=0$\cite{Cuccoli_PRB_2003}. The four experimental values are based on the information given in Table~\ref{comparison-table} with $\Delta$ = $H_{\mathrm{A}}/H_{\mathrm{SAT}}$. The error bars of the data are smaller than the sizes of the symbol.}
\label{cuccoliplot}
\end{center}
\end{figure}

This equation is plotted as the solid line in Fig.~\ref{cuccoliplot}. It is based on the assumption that $J^\prime = 0$; the presence of a finite $J^\prime$ would raise the curve. By putting in the values of the exchange anisotropy for the three copper/pyrazine compounds as determined by the $H_{\mathrm{A}}/H_{\mathrm{SAT}}$, Table~\ref{comparison-table}, Eq.~(\ref{cuccolieqn}) can be used to predict the critical ratios for the three compounds, in the absence of the  $J^\prime$  interaction. For the ClO$_4$, BF$_4$, and PF$_6$ compounds, the critical ratios would be 0.20, 0.22, and 0.16, respectively. The corresponding value for Sr$_2$CuO$_2$Cl$_2$ is 0.15. The experimental critical ratios of the same four compounds are also plotted on Fig.~\ref{cuccoliplot}.  In this plot, if the transition is primarily driven by the $XY$ anisotropy, one anticipates the experimental data will be found somewhat above the curve, the exact amount determined by $J^\prime$. At the opposite limit, with the transition primarily driven by the interlayer interactions, the experimental value will appear substantially above the curve.

Fig.~\ref{cuccoliplot} shows three of the compounds follow a similar pattern while the fourth is a definite outlier. The values for ClO$_4$, BF$_4$, and Sr$_2$CuO$_2$Cl$_2$ follow the general shape of the curve but lie above it by 0.04 (0.24 - 0.20) for ClO$_4$, 0.02 units for BF$_4$ and 0.03 for Sr$_2$CuO$_2$Cl$_2$. This behavior is consistent with their description as quasi-2D systems in which the stronger perturbation to the basic 2D QHAF Hamiltonian is the moderate ($\sim 0.5$\% for ClO$_4$ and BF$_4$) $XY$ anisotropy. In the absence of interlayer interactions, the critical ratios for these three compounds would have been been located on the solid curve but the presence of $J^\prime$ enhanced the transition temperature approximately 20\% to their measured values. Since their ordering is primarily due to the anisotropy, the values for $J^\prime/J$ in Table~\ref{comparison-table} ($8 \times 10^{-4}$ and $1 \times 10^{-3}$, respectively) derived from Eq.~(\ref{Yasuda}) (which assumed $\Delta=0$) are clearly too high. The appropriate values of $J^\prime/J$ are those which will raise the critical ratio of 0.20 for a purely 2D anisotropic magnet to 0.24 for a {\em quasi-}2D anisotropic magnet. In the absence of guidance by simulations, it is not useful to speculate about the appropriate values. We do note that the closer proximity of the BF$_4$ value to the theoretical curve means that it has better isolation between the layers than does the  ClO$_4$ compound.

In contrast, the value for [Cu(pz)$_2$(NO$_3$)](PF$_6$) is found 0.13 units higher than the BKT curve. Although it has an $XY$ anisotropy smaller  by a factor of thirty than those of the ClO$_4$ and BF$_4$ compounds, it has the highest critical ratio of the three copper/pyrazine compound. It is clear that its transition is driven by the 3D interactions. Consequently, its $J^\prime/J$ ratio of $3 \times 10^{-3}$, as derived from Eq.~(\ref{Yasuda}), is a good approximation. [Cu(pz)$_2$(NO$_3$)](PF$_6$) is therefore a remarkably isotropic quasi-Heisenberg copper compound but the least well-isolated of the three compounds in this study.

\subsection{Field-Induced Anisotropy}
In addition to interlayer interactions and intrinsic exchange anisotropy, there is a third contribution to the Hamiltonian (Eq.~(\ref{Hamiltonian})) that can induce an ordering transition in a 2D QHAF, the magnetic field. It has been recognized for over a decade\cite{TroyerPRL1998} that an external field is equivalent to an easy-plane anisotropy in its ability to induce BKT transitions in layered Heisenberg antiferromagnets. This work was later extended~\cite{Cuccoli_PRB_2003} to show that an external field could induce a minimum in the susceptibility of an isotropic ($\Delta$= 0) 2D $S=1/2$ antiferromagnet. In the case of an $XY$ magnet, the field would enhance the minimum in $\chi_z$ and overcome the intrinsic anisotropy to induce a minimum in $\chi_x$ for sufficiently strong fields.
	
The effects of field-induced anisotropy on the ordering temperature in a quasi-2D QHAF was first reported in the 1995 study of Sr$_2$CuO$_2$Cl$_2$ single crystals by NMR and susceptibility~\cite{Suh1995PRL}. The zero-field $T_{\mathrm{N}}$ for this compound is 256.5~K. When the field was applied in the easy plane, $T_{\mathrm{N}}$ increased by 2.3\% to 262.5~K in a 4.7~T field, and by 3.4\% to 265.3~K in an 8.2~T field; when the field was applied normal to the planes, no enhancement in $T_{\mathrm{N}}$ was observed.
	
More dramatic field-induced effects have been observed in the ClO$_4$, BF$_4$, and PF$_6$ compounds, made possible by their exchange strengths being only $\sim 1\%$ as large as that for Sr$_2$CuO$_2$Cl$_2$. For external fields exceeding the respective anisotropy fields (Table~\ref{comparison-table}), minima occur for the $\chi_x$ of each of the compounds. Evidence of this effect is found in Fig.~\ref{tob-chi-both} for PF$_6$. These data were collected in a 1~kOe field, much stronger than the 70~Oe anisotropy field for this compound. As revealed by the low-field ac-susceptibility measurements (Fig.~\ref{tob-chi-both}) only when the dc-field was reduced to less than 70 Oe did the minimum in $\chi_x$ vanish. In addition, the N\'{e}el temperatures of all compounds strongly increased in the presence of larger fields. For the ClO$_4$ compound, $T_{\mathrm{N}}$ rose from 4.25~K in zero-field to 5.7~K in a 15 T field, an increase of 34\%. Unexpectedly, the field-induced enhancement of $T_{\mathrm{N}}$ was the same independent of the orientation of the field.
	A detailed study of these effects and a discussion of the magnetic phase diagram are in progress and will be reported in a subsequent publication.

\section{SUMMARY}
The magnetic properties of three molecular-based quasi-2d S=1/2 Heisenberg antiferromagnets ((Cu(pz)$_2$(ClO$_4$)$_2$, Cu(pz)$_2$(BF$_4$)$_2$, and (Cu(pz)$_2$(ClO$_4$)$_2$ have been investigated. The ordering temperatures have been determined both by  $\mu^{+}$SR and susceptibility studies. Values of the intralayer and interlayer exchange strengths ($J$,  $J^\prime$) and  the exchange-anisotropy parameters $\Delta$ have been determined. Cu(pz)$_2$(ClO$_4$)$_2$ and Cu(pz)$_2$(BF$_4$)$_2$ have similar exchange strengths (17.5 and 15.3 K, respectively), similar critical ratios $T_{\mathrm{N}}/J$ (0.243 and 0.248, respectively), and similar ratios of anisotropy fields to saturation fields (5.3$\times$10$^{-3}$ and 5.8$\times$10$^{-3}$, respectively). [Cu(pz)$_2$(NO$_3$)](PF$_6$) has a a weaker exchange strength, a much smaller anisotropy field, but the highest critical temperature; it is the least 2D but the most Heisenberg-like of the three compounds. The experimental data are all explained by the existence of small $XY$ contributions to the spin Hamiltonian, as well as weak coupling between the magnetic layers. The influence of field-induced anisotropy on the magnetic behavior has been observed.

\section{Acknowledgments}
The authors thank Matthias Troyer (ETH) for the use of his simulations of the magnetization of the 2D QHAF and Mike Zhitomirsky for sharing with us the numerical results of his spin-wave magnetization calculation. The Clark University Quantum Design MPMS-XL magnetometer was purchased with the assistance of the NSF (through Grant IMR-0314773) and the Kreske Foundation. Part of this work was carried out at the National High Magnetic Field Laboratory which is supported by the National Science Foundation Cooperative Agreement No. DMR 0654118 and by the State of Florida.
$\mu^{+}$SR measurements were carried out at the ISIS Facility and the Swiss Muon Source. We are grateful to the EPSRC (UK) for
financial support and to Alex Amato for experimental assistance in making these measurements. We also thank Jamie Manson whose earlier muon study of the PF$_6$ compound\cite{Lancaster07prl} helped us to understand our own data.

\end{document}